\documentclass[a4,aps,prd,superscriptaddress,floatfix,nofootinbib,showpacs]{revtex4}

\usepackage{color}
\usepackage[dvips]{graphicx}

\newcommand{\lsim}{\mathrel{\mathop{\kern 0pt \rlap
  {\raise.2ex\hbox{$<$}}}
  \lower.9ex\hbox{\kern-.190em $\sim$}}}
\newcommand{\gsim}{\mathrel{\mathop{\kern 0pt \rlap
  {\raise.2ex\hbox{$>$}}}
  \lower.9ex\hbox{\kern-.190em $\sim$}}}

\newcommand{\be}{\begin{equation}}
\newcommand{\ee}{\end{equation}}
\newcommand{\beqarr}{\begin{eqnarray}}
\newcommand{\eeqarr}{\end{eqnarray}}

\begin{document}

%\preprint{DFTT 42/2002}

\title{Light Relic Neutralinos}
\thanks{Preprint number: DFTT 42/2002}

% address or url should go in the {}'s for \email and \homepage.
% Please use the appropriate macro for each each type of information

% \affiliation command applies to all authors since the last
% \affiliation command. The \affiliation command should follow the
% other information
% \affiliation can be followed by \email, \homepage, \thanks as well.
  
%
\author{A. Bottino}
\email{bottino@to.infn.it}
\homepage{http://www.to.infn.it/astropart}
\affiliation{Dipartimento di Fisica Teorica, Universit\`a di Torino \\
Istituto Nazionale di Fisica Nucleare, Sezione di Torino \\
via P. Giuria 1, I--10125 Torino, Italy}

\author{N. Fornengo}
\email{fornengo@to.infn.it}
\homepage{http://www.to.infn.it/astropart}
\homepage{http://www.to.infn.it/~fornengo}
\affiliation{Dipartimento di Fisica Teorica, Universit\`a di Torino \\
Istituto Nazionale di Fisica Nucleare, Sezione di Torino \\
via P. Giuria 1, I--10125 Torino, Italy}

\author{S. Scopel} 
\email{scopel@to.infn.it}
\homepage{http://www.to.infn.it/astropart}
\affiliation{Dipartimento di Fisica Teorica, Universit\`a di Torino \\
Istituto Nazionale di Fisica Nucleare, Sezione di Torino \\
via P. Giuria 1, I--10125 Torino, Italy}

\date{\today}

\begin{abstract} \vspace{1cm}  
The relic abundance and the scalar cross--section off nucleon for light
neutralinos (of mass below about 45 GeV) are evaluated in an effective
MSSM model without GUT--inspired relations among gaugino masses. It is
shown that these neutralinos may provide a sizeable contribution to
the matter density in the Universe and produce measurable effects in
WIMP direct detection experiments. These properties are elucidated in
terms of simple analytical arguments.
\end{abstract}

\pacs{95.35.+d,11.30.Pb,12.60.Jv,95.30.Cq}
% 11.30.Pb Supersymmetry
% 12.60.Jv Supersymmetric models
% 95.30.Cq Elementary particle processes
% 95.35.+d Dark matter
% 98.35.Gi Galactic halo (Milky Way)
% 98.35.Df Kinematics, dynamics, and rotation
% 98.35.Pr Solar neighborhood

\maketitle

\section{Introduction}
\label{sec:intro}

Most works on relic neutralinos consider supersymmetric schemes with a
unification assumption for the gaugino masses $M_i$ ($i = 1,2,3$) at
the GUT scale $M_{GUT} \sim 10^{16}$ GeV. This hypothesis implies that
at lower scales the following relations hold:
\begin{equation}
M_1 : M_2 : M_3 = \alpha_1 : \alpha_2 : \alpha_3, 
\label{sec:unif}
\end{equation}
where the $\alpha_i$ ($i = 1,2,3$) are the coupling constants of the
three Standard Model gauge groups. In particular, at the electroweak
scale, $M_{EW} \sim 100$ GeV, $M_1$ and $M_2$ are related by the
expression:
\begin{equation}
M_1 =  \frac{5}{3} \tan^2 \theta_{W} \; M_2 \simeq 0.5~M_2. 
\label{eq:gut}
\end{equation}
However, there are theoretical arguments for considering
supersymmetric schemes where the unification assumption on gaugino
masses is not satisfied\cite{nonunif}.

In the present paper we analyse properties of relic neutralinos in an
effective Minimal Supersymmetric extension of the Standard Model
(MSSM) where the GUT relation of Eq. (\ref{eq:gut}) is relaxed.
Previous papers where supersymmetric schemes without gaugino masses
unification have been considered in connection with relic neutralinos
include the ones reported in Refs.
\cite{dt,gr,mny,go,bbd,cn,ads,cerdeno,bk,bbb,bno,bhn}. Here we
evaluate the neutralino relic abundance $\Omega_{\chi} h^2$ and the
neutralino-nucleon scalar cross-section $\sigma_{\rm scalar}^{(\rm
nucleon)}$, which is relevant to dark matter direct detection. In
Sect.\ref{sec:model} we define the supersymmetric scheme adopted in
the present paper and in Sect. \ref{sec:cross} we provide analytical
considerations and numerical evaluations. Our conclusions are drawn in
Sect. \ref{sec:finale}.

\section{Effective MSSM without gaugino unification} 
\label{sec:model}

We employ an effective MSSM scheme (effMSSM) at the electroweak scale,
defined in terms of a minimal number of parameters, only those
necessary to shape the essentials of the theoretical structure of MSSM
and of its particle content.  The assumptions that we impose at the
electroweak scale are: a) all squark soft--mass parameters are taken
degenerate: $m_{\tilde q_i} \equiv m_{\tilde q}$; b) all slepton
soft--mass parameters are taken degenerate: $m_{\tilde l_i} \equiv
m_{\tilde l}$;  c) all trilinear parameters are set to zero except
those of the third family, which are defined in terms of a common
dimensionless parameter $A$: $A_{\tilde b} = A_{\tilde t} \equiv A
m_{\tilde q}$ and $A_{\tilde \tau} \equiv A m_{\tilde l}$.  As a
consequence, the supersymmetric parameter space consists of the
following independent parameters: $M_2, \mu, \tan\beta, m_A, m_{\tilde
q}, m_{\tilde l}, A$ and $R \equiv M_1/M_2$.  In the previous list of
parameters we have denoted by $\mu$ the Higgs mixing mass parameter,
by $\tan\beta$ the ratio of the two Higgs v.e.v.'s and by $m_A$ the
mass of the CP-odd neutral Higgs boson.

This scheme differs from the effMSSM which we employed for instance in
Ref. \cite{pinningetc} in the fact that we are relaxing here the
gaugino unification relation, which was instead assumed in our
previous works. The presence of the extra $R$ parameter accounts for
this fact.

The neutralino is defined as the lowest--mass linear superposition of
bino $\tilde B$, wino $\tilde W^{(3)}$ and of the two higgsino states
$\tilde H_1^{\circ}$, $\tilde H_2^{\circ}$:
\begin{equation}
\chi \equiv a_1 \tilde B +
a_2 \tilde W^{(3)} + a_3 \tilde H_1^{\circ} + a_4 \tilde H_2^{\circ}.
\end{equation}
Due to well-known properties of the neutralino and chargino mass
matrices, one has that: a) for $\mu \gg M_1, M_2$ the neutralino mass
is determined by the lightest gaugino mass parameter: $m_{\chi} \simeq
\min(M_1,M_2)$, while the lightest chargino mass is set by $M_2$:
$m_{\chi^{\pm}} \simeq M_2$ ($M_1$ does not enter the chargino mass
matrix at tree--level); b) for $\mu \ll M_1, M_2$ both the neutralino
and the chargino masses are primarily set by the Higgs mixing
parameter: $m_{\chi} \simeq \mu \simeq m_{\chi^{\pm}}$.

LEP data put a stringent lower bound on the chargino mass:
$m_{\chi^{\pm}} \gsim 103 $ GeV, which converts into lower bounds on
$M_2$ and $\mu$: $M_2, \mu \gsim $ 103 GeV. This implies a lower bound
on the neutralino mass of the order of about 50 GeV in the standard
effMSSM, where the GUT relation of Eq.(\ref{eq:gut}) holds. On the
contrary, the neutralino mass may be smaller when $M_1 \ll M_2$, thus
for small values of the parameter $R$.

In the present paper we are interested in the phenomenology of light
neutralinos, therefore we consider values of $R$ lower than its GUT
value: $R_{GUT} \simeq 0.5$. For definiteness we will consider
the range: 0.01 - 0.5. The ensuing light neutralinos have a dominant
bino component; a deviation from a pure bino composition is mainly
due to a mixture of $\tilde B$ with $\tilde H_1^{\circ}$, as will
be shown in Sect.\ref{sec:results}.

In our numerical analysis we have varied the MSSM parameters within
the following ranges: $1 \leq \tan \beta \leq 50$, $100\, {\rm GeV
}\leq |\mu|, M_2, m_{\tilde q}, m_{\tilde l} \leq 1000\, {\rm GeV }$,
${\rm sign}(\mu)=-1,1$, $90\, {\rm GeV }\leq m_A \leq 1000\, {\rm GeV
}$, $-3 \leq A \leq 3$, for a sample of representative values of $R$
in the range $0.01 \leq R \leq 0.5$. This range for $R$, implemented
with the experimental lower limit on $M_2$ of about 100 GeV, implies
that the lower bound on the neutralino mass can be moved down to few
GeV's for $R\sim 0.01$.

We then implemented the following experimental constraints:
accelerators data on supersymmetric and Higgs boson searches (CERN
$e^+ e^-$ collider LEP2 \cite{LEPb} and Collider Detector CDF at
Fermilab \cite{cdf}); measurements of the $b \rightarrow s + \gamma$
decay \cite{bsgamma}. We wish to comment that the accelerator limits
on the Higgs sector are taken into account by implementing the limits
on the Higgs production cross--sections: $e^+ e^- \rightarrow h Z$ and
$e^+ e^- \rightarrow h A$ ($h$ and $A$ are the lightest scalar and the
pseudoscalar neutral Higgs bosons, respectively), which in turn imply
a constraint on the coupling constants $\sin^2(\alpha-\beta)$ and
$\cos^2(\alpha-\beta)$. Once these limits are applied, the absolute
lower limit on the Higgs masses is $m_A,m_h \sim 90$ GeV. The allowed
light--Higgs mass range between 90 and 114 GeV is very often
overlooked in studies of neutralino dark matter, where a flat limit of
114 GeV is applied to $m_h$. The light--Higgs mass range, even though
difficult (but not impossible) to be achieved in SUGRA models
\cite{higgs,prob,weiglein}, is nevertheless quite natural in the
effMSSM and usually provides large detection rates for neutralino dark
matter \cite{prob}.

As for the constraint due to the muon anomalous magnetic moment $a_\mu
\equiv (g_{\mu} - 2)/2$ we have used the interval $-160 \leq \Delta
a_{\mu} \cdot 10^{11} \leq 680 $, where $\Delta a_{\mu}$ is the
deviation of the current world average of the experimental
determinations (dominated by the measurements of Ref. \cite{amm}) from
the theoretical evaluation within the Standard Model: $\Delta a_{\mu}
\equiv a_{\mu}^{\rm expt} - a_{\mu}^{SM}$.  The range we use for
$\Delta a_{\mu}$ is a 2$\sigma$ interval, obtained by using for the
lowest-order hadronic vacuum polarization contribution an average
between the results derived from the $e^+ - e^-$ data \cite{dehz,hmnt}
and from hadronic $\tau$ decays \cite{dehz}. The $\Delta a_\mu$
constraint and the $b \rightarrow s + \gamma$ bound set stringent
limits for the light neutralino sector of our models.

Once also the relic abundance bound $\Omega_\chi h^2 \leq 0.3 $ is
applied (see Sect.\ref{sec:results}) in addition to the other
experimental constraints discussed above, a lower limit of about 6 GeV
is obtained for the neutralino mass in the class of models with
non--universal gaugino masses considered in this paper\footnote{This
is at variance with the results of Ref. \cite{boudjema}, where a lower
limit on the neutralino mass of 12 GeV has been deduced. Notice that
the authors of Ref. \cite{boudjema} consider only the limit of very
large $m_A$, which strongly suppresses processes which involve
$A$--exchange, in particular the neutralino annihilation cross
section. On the contrary, we are considering also the light Higgs
sector, which is effective in reducing the value of the neutralino
relic abundance and therefore in allowing lighter neutralinos (see
Eq.(\ref{eq:omega1})).}.

\section{Neutralino relic abundance and neutralino-nucleon cross-section} 
\label{sec:cross}

\subsection{Some analytical properties for small $m_{\chi}$}
\label{sec:analytic}

The neutralino configurations which provide the highest values of 
direct detection rates are the ones dominated by $(h,H)$ Higgs-exchange 
processes, which in turn require a gaugino-higgsino mixing. 
For these configurations, also the relic abundance is regulated by a  
($A$)Higgs-exchange diagram in the $\chi-\chi$ annihilation cross-section. 

Thus, to get an insight into the properties to be expected for our
light neutralinos we limit ourselves to the following approximate
expressions, derived under the assumptions of Higgs-dominance and
light neutralinos (notice however that full exact expressions both for
the relic abundance $\Omega_{\chi} h^2$ and for the
neutralino--nucleon scalar cross section $\sigma_{\rm scalar}^{(\rm
nucleon)}$ are employed in the numerical evaluations to be discussed
in the next Section). Under these hypotheses, the neutralino relic
abundance is dominated by the s--wave annihilation in a $\bar b b$
pair (unless $m_\chi$ is very close to the $b$--quark mass $m_b$, in
which case the $\bar c c$ and $\bar \tau \tau$ channel are dominant):
\begin{equation}
\Omega_{\chi} h^2 \simeq \frac{4 \cdot 10^{-39} {\rm
cm}^2}{\langle\sigma_{\rm ann} \; v\rangle_{\rm int}} \simeq
\frac{10^{-37} {\rm cm}^2}{6 \pi \alpha_{em}^2}
\frac{\sin^4\theta_{W}}{ \tan^2\beta (1+\epsilon)^2} (a_2 - a_1 \tan
\theta_{W})^{-2} (a_4 \cos \beta - a_3 \sin \beta)^{-2} \frac{[(2
m_{\chi})^2 - m_A^2]^2}{m_{\chi}^2~[1-m_b^2/m_\chi^2]^{1/2}} \frac{m_W^2}{m_b^2}\, ,
\label{eq:omega0}
\end{equation}
and the elastic scattering cross section is:
\begin{equation}
\sigma_{\rm scalar}^{(\rm nucleon)} \simeq \frac {8 G_F^2} {\pi} M_Z^2 
m_{\rm red}^2 \; 
\left[\frac{F_h I_h}{m_h^2}+\frac{F_H I_H}{m_H^2} \right]^2 \, .
\label{eq:sigmasc}
\end{equation}
In the previous equations we have used the following notations:
${\langle \sigma_{\rm ann} \; v\rangle_{\rm int}}$ is the integral
from present temperature up to the freeze-out temperature of the
thermally averaged product of the annihilation cross-section times the
relative velocity of a pair of neutralinos; $\epsilon$ is a quantity
which enters in the relationship between the down--type fermion
running masses and the corresponding Yukawa couplings (see, for
instance, Refs. \cite{size,higgs} and references quoted therein);
$m_{\rm red}$ is the neutralino-nucleon reduced mass. The quantities
$F_{h,H}$ and $I_{h,H}$ are defined as follows:
\beqarr F_h &=& (-a_1 \sin \theta_W+a_2 \cos \theta_W) (a_3
\sin \alpha + a_4 \cos \alpha)\, , \nonumber 
\\ 
F_H &=& (-a_1 \sin
\theta_W+a_2 \cos \theta_W) (a_3 \cos \alpha - a_4 \sin \alpha)\, ,
\nonumber 
\\ 
I_{h,H}&=&\sum_q k_q^{h,H} m_q \langle N|\bar{q} q |N
\rangle \, .
\label{eq:effe}
\eeqarr
The matrix elements $<N|\bar{q}q|N>$ are meant over the nucleonic
state. The values adopted here for $m_q <N|\bar{q}q|N>$ are the ones
denoted by set 1 in Ref. \cite{implications}. We remind that
uncertainties in the values of $m_q <N|\bar{q}q|N>$ can give rise to
an increase of the neutralino--nucleon cross section of about a factor
of a few \cite{size}.

The angle $\alpha$ rotates $H_1^{(0)}$ and $H_2^{(0)}$ into $h$ and
$H$, and the coefficients $k_q^{h,H}$ are given by

\begin{eqnarray}
k_{u{\rm -type}}^h &=& ~\cos\alpha / \sin\beta  \, , \nonumber\\
k_{d{\rm -type}}^h &=& - \sin\alpha / \cos\beta -\epsilon \cos(\alpha-\beta) \tan\beta \, , \nonumber\\
k_{u{\rm -type}}^H &=& ~\sin\alpha / \sin\beta \, , \nonumber\\
k_{d{\rm -type}}^H &=& ~\cos\alpha / \cos\beta -\epsilon \sin(\alpha-\beta) \tan\beta \, , 
\label{eq:k}
\end{eqnarray}
for the up--type and down--type quarks, respectively. 

In the discussion which follows we only wish to establish some
correlations implied by the dependence of $\Omega_{\chi} h^2$ and of
$\sigma_{\rm scalar}^{(\rm nucleon)}$ on the Higgs masses and the neutralino
mass. For this purpose we rewrite the two previous expressions as
follows:
\begin{equation}
\Omega_{\chi} h^2 \simeq C \;
\frac{[(2 m_{\chi})^2 - m_A^2]^2}{m_{\chi}^2~[1-m_b^2/m_\chi^2]^{1/2}} \, ,
\label{eq:omega}
\end{equation}
\begin{equation}
\sigma_{\rm scalar}^{(\rm nucleon)} \simeq \frac{D}{m_h^4} \, ,
\label{eq:sigma}
\end{equation}
with obvious definitions for $C$ and $D$. Here $m_h$ stands
generically for the mass of the one of the two CP-even neutral Higgs
bosons which provides the dominant contribution to
$\sigma_{\rm scalar}^{(\rm nucleon)}$.

We now consider the case of very light neutralinos, 
{\it i.e.} $m_{\chi} \ll \frac{1}{2} m_A$. Therefore we may further 
simplify Eq. (\ref{eq:omega}) as 
\begin{equation}
\Omega_{\chi} h^2 \simeq C 
\frac{m_A^4}{m_{\chi}^2~[1-m_b^2/m_\chi^2]^{1/2}} 
\label{eq:omega1}
\end{equation}

The largest neutralino--nucleon scattering cross sections occur when
both $m_h$ and $m_A$ are close to their experimental lower bound
($m_h\sim m_A \sim 90-100$ GeV) and $\tan\beta$ is relatively large,
in which case also the couplings of Eqs. (\ref{eq:effe},\ref{eq:k})
between neutralinos and down--type quarks through $h$--exchange are
sizeable \cite{higgs}. In this case, from Eqs. (\ref{eq:sigma}) and
(\ref{eq:omega1}) one derives the range of $\sigma_{\rm scalar}^{(\rm
nucleon)}$ at fixed value of $m_{\chi}$ (always in the regime
$m_{\chi} \ll \frac{1}{2} m_A$):
\begin{equation}
\frac{C \; D}{m_{\chi}^2~[1-m_b^2/m_\chi^2]^{1/2}~(\Omega_{\chi}
h^2)_{\rm max}} \lsim \sigma_{\rm scalar}^{(\rm nucleon)} \lsim
\frac{D}{m_{h,\rm min}^4},
\label{eq:range}
\end{equation}

\noindent
where $m_{h,min}$ stands for the experimental lower bound on $m_h$.
The lower limit to $\sigma_{\rm scalar}^{(\rm nucleon)}$ displayed in
Eq. (\ref{eq:range}) provides a stringent lower bound on $\sigma_{\rm
scalar}^{(\rm nucleon)}$ for very light neutralinos. This feature will
show up in the numerical evaluations presented in the next Section.
The upper bound on $\sigma_{\rm scalar}^{(\rm nucleon)}$ is instead
determined by the lower limit on the Higgs mass $m_h$.

By the arguments given above, it turns out that in the small mass
regime ($m_\chi \ll \frac{1}{2} m_A$) the upper bound on the relic
abundance $\Omega_{\chi} h^2 \leq 0.3$ establishes a constraint
between the otherwise independent parameters $m_{\chi}$ and $m_A$
(see Eq.(\ref{eq:omega1})).

\subsection{Numerical results}
\label{sec:results}

We turn now to the presentation of our numerical results. In
Figs.\ref{fig:1a}--\ref{fig:1b} we give the scatter plots of the
quantity $\xi \sigma_{\rm scalar}^{(\rm nucleon)}$ in terms of the
neutralino mass for different values of the parameter $R$. The
quantity $\xi$ is defined as the ratio of the local (solar
neighbourhood) neutralino matter density to the total local dark
matter density: $\xi \equiv \rho_{\chi}/\rho_{\rm loc}$. In
Figs.\ref{fig:1a}--\ref{fig:1b} we plot the quantity $\xi \sigma_{\rm
scalar}^{(\rm nucleon)}$, rather than simply $\sigma_{\rm
scalar}^{(\rm nucleon)}$, in order to include in our considerations
also neutralino configurations of low relic abundance ({\it i.e.}
cosmologically subdominant neutralinos). We recall that, from
experimental measurements of the direct detection rates, only the
product $\xi \sigma_{\rm scalar}^{(\rm nucleon)}$ may be extracted,
and not directly $\sigma_{\rm scalar}^{(\rm nucleon)}$. The quantity
$\xi$ is derived here from the relic abundance by the usual rescaling
recipe: $\xi = {\rm min}(1,\Omega_\chi h^2/[\Omega h^2]_{\rm min})$,
where the minimal value of relic abundance which defines a neutralino
as a dominant dark matter component has been fixed at the value
$[\Omega h^2]_{\rm min} = 0.05$. $\Omega_\chi h^2$ and $\sigma_{\rm
scalar}^{(\rm nucleon)}$ are evaluated according to the procedures and
formulae described in Refs. \cite{relic,implications}.

Figs.\ref{fig:1a}--\ref{fig:1b} displays quite remarkable properties
of the light relic neutralinos from the point of view of their
detectability by WIMP direct measurements. These properties are easily
understandable in terms of the analytic arguments presented in the
previous Section. For instance, in each panel at a fixed value of $R
\lsim 0.1$, there is a characteristic funnel pointing toward high
values of $\xi \sigma_{\rm scalar}^{(\rm nucleon)}$ at small
neutralino masses. This originates in the lower bound on $\sigma_{\rm
scalar}^{(\rm nucleon)}$ reported in Eq.(\ref{eq:range}), which is
effective only for very low neutralino masses (below about 15 GeV) and
becomes more and more stringent as $m_{\chi}$ decreases. As displayed
in Eq.(\ref{eq:range}), the size of this lower bound, apart from
relevant supersymmetric details, is determined by the value of
$(\Omega_\chi h^2)_{\rm max}$, which is set here at the value
$(\Omega_\chi h^2)_{\rm max} = 0.3$. It is noticeable that at very
small values of $R$, for instance at $R = 0.01$, all supersymmetric
configurations are within the cosmologically interesting range of
$\Omega_{\chi}$ ({\it i.e.} no configuration of this set is rescaled)
and provide large values of $\sigma_{\rm scalar}^{(\rm nucleon)}$
({\it i.e.} large detection rates).

As we increase the value of $R$, in our scan we are accessing larger
values of $m_\chi$: again the largest values of $\xi \sigma_{\rm
scalar}^{(\rm nucleon)}$ are dominated by Higgs--exchange, for Higgs
masses close to their lower bound of about 90 GeV. This is also true
for the annihilation cross section. This approaches its pole at
$m_\chi \sim m_A/2$; therefore, the largest values of $\xi \sigma_{\rm
scalar}^{(\rm nucleon)}$ refer to subdominant neutralinos, as $m_\chi$
increases toward $m_\chi \sim 45$ GeV (which represents the pole in
the annihilation cross section for the lightest possible $A$
boson). These features are clearly shown in
Figs.\ref{fig:1a}--\ref{fig:1b}. The panel denoted by ``standard'' in
Fig.\ref{fig:1b} refers to the usual case of universal gaugino masses:
in this case the neutralino mass is bounded from below at about 50
GeV, and therefore all the interesting low neutralino--mass sector is
precluded.  The last panel in Fig.\ref{fig:1b} (denoted by ``global'')
shows our results for $R$ varied in the interval $0.01-0.5$: the
funnel at low masses and the effect of the $A$--pole in the
annihilation cross section are clearly visible.

We recall that, for each panel at fixed $R$, the lower value of the
neutralino mass is a consequence of the experimental bound on the
chargino mass, which in turn fixes a lower bound on $M_1= R \times
M_2$. The upper value on the neutralino mass for each panel is a mere
consequence of the fact that we scan the $M_2$ parameter up to 1 TeV.

The detailed connection among the values of $\sigma_{\rm scalar}^{(\rm
nucleon)}$ and those of $\Omega_{\chi} h^2$ is given in
Fig. \ref{fig:2}.  The strong correlation between $\sigma_{\rm
scalar}^{(\rm nucleon)}$ and $\Omega_{\chi} h^2$ displayed for $R =
0.01$ reflects the properties of the funnel previously discussed in
connection with Fig.\ref{fig:1a}. All the configurations refer to
large values of $\Omega_\chi h^2$: actually, it is the upper bound on
the neutralino relic abundance which determines the strong bound on
the allowed configurations. By changing $R$ from 0.01 to larger
values, we observe that the ensuing increase in $m_\chi$ shifts the
configurations of largest $\sigma_{\rm scalar}^{(\rm nucleon)}$ toward
lower values of relic abundance, as expected from the analitical
considerations of the previous Section. From this figure we see that a
fraction of the largest values of the quantity $\xi \sigma_{\rm
scalar}^{(\rm nucleon)}$ refer to dominant neutralinos, while another
fraction refers to slightly subdominant neutralinos: $0.01 \lsim
\Omega_\chi h^2 \lsim 0.05$. Configurations with $\Omega_\chi h^2 <
0.01$, even providing the largest values of the scattering cross
section (see, for instance, the panel at $R=0.04$ in Fig. \ref{fig:2})
suffer from a severe rescaling factor $\xi$ which somehow reduces
their detectability.

The fact that for small values of $R$ the scattering and
neutralino--neutralino annihilation cross sections are dominated by
Higgs exchange is a consequence of two facts: the relatively small
values for the lower bounds on $m_h$ and $m_A$ and the neutralino
composition, which, even though dominated by the bino component,
nevertheless possesses a non negligible higgsino contribution allowing
the neutralino to efficiently couple with the Higgs fields.

Fig. \ref{fig:3} shows that for small values of $R$ (small $m_{\chi}$)
the neutralino-neutralino annihilation cross-section is indeed
dominated by Higgs-exchange diagrams, especially for the largest
values of $\sigma_{\rm scalar}^{(\rm nucleon)}$. The first panel of
Fig. \ref{fig:3}, which refers to $R=0.01$, clearly shows that the
annihilation cross section is strongly dominated by Higgs exchange.
For $R=0.02$ the annihilation cross section can be either dominated by
Higgs or sfermion exchange: however, the configurations which provide
values of $\sigma_{\rm scalar}^{(\rm nucleon)}$ in excess of $10^{-8}$
nbarn (denoted by crosses) show a clear Higgs dominance in the
annihilation cross section. These features are progressively lost when
$R$ increases: the annihilation cross section may be dominated by $Z$
exchange (which, by coincidence, has its pole also at about 45 GeV).

Finally, Fig. \ref{fig:4} shows that for low values of $R$, the
neutralino composition is dominated by the bino component, but a
deviation from a pure bino composition is present and is mainly due to
a mixture of $\tilde B$ with $\tilde H_1^0$. The two composition
parameters $a_1^2$ and $a_3^2$ remain aligned along the $a_1^2+a_3^2 =
1$ diagonal line up to $R\sim 0.05$, with a clear dominance (above
$70\%$) in bino. For larger values of $R$ the correlation between
$a_1^2$ and $a_3^2$ starts to deviate from the diagonal line, a fact
that indicates how the two other components are becoming important (it
is mainly $a_4$ which sets up). The panel at $R=0.1$ shows that the
bino component is usually large, but a sizeable mixture starts
occurring.  The last panel in Fig. \ref{fig:4} recalls the situation
for the standard case of universal gaugino masses, where the
neutralino may be any mixture of its component fields.

\section{Conclusions}
\label{sec:finale}

In the present paper we have focussed our attention on relic
neutralinos of light masses: $m_{\chi} \lsim $ 45 GeV, which are
allowed in supersymmetric models where no unification of gaugino
masses is assumed. We have shown that these neutralinos may have
elastic cross-sections off nucleons which go up to $\sigma_{\rm
scalar}^{(\rm nucleon)} \sim 10^{-7}$ nbarn, with a relic abundance of
cosmological interest: $0.05 \lsim \Omega_{\chi} h^2 \lsim 0.3$.

The present upper limits to $\xi \; \sigma_{\rm scalar}^{(\rm
nucleon)}$ provided by WIMP direct detection experiments
\cite{morales,dama0,cdms,edel} do not constrain the supersymmetric
configurations for the light neutralinos considered here. This is
especially true once the relevant uncertainties (mainly related to the
form and parameters of the WIMP galactic distribution function
\cite{bcfs} and to the quenching factors for bolometric detectors) are
taken into account. The CDMS upper bound \cite{cdms} could concern a
small fraction of supersymmetric configurations in the range around 15
GeV, though very marginally, if the uncertainties on astrophysical
quantities are considered. Moreover, the CDMS bound needs a
confirmation by a further running in a deep--underground site, as
planned by the Collaboration.

The small--mass neutralino configurations analysed in the present
paper are accessible to experiments of direct detection with a
low--energy threshold and a high sensitivity.  An experiment of this
type is the DAMA experiment with a mass of $\simeq$ 100 kg of NaI(Tl),
whose results after a 4-years running show an annual-modulation effect
at a $4\sigma$ C.L. which does not appear to be related to any
possible source of systematics \cite{damapl}. The DAMA experiment,
with its high sensitivity, is potentially good to investigate also the
relic neutralinos considered in the present paper.

\begin{acknowledgements}
We gratefully acknowledge financial support provided by Research
Grants of the Italian Ministero dell'Istruzione, dell'Universit\`a e
della Ricerca (MIUR) and of the Universit\`a di Torino within the {\sl
Astroparticle Physics Project}.
\end{acknowledgements}

\newpage

\renewcommand{\thefigure}{1\alph{figure}}
\begin{figure} \centering
\includegraphics{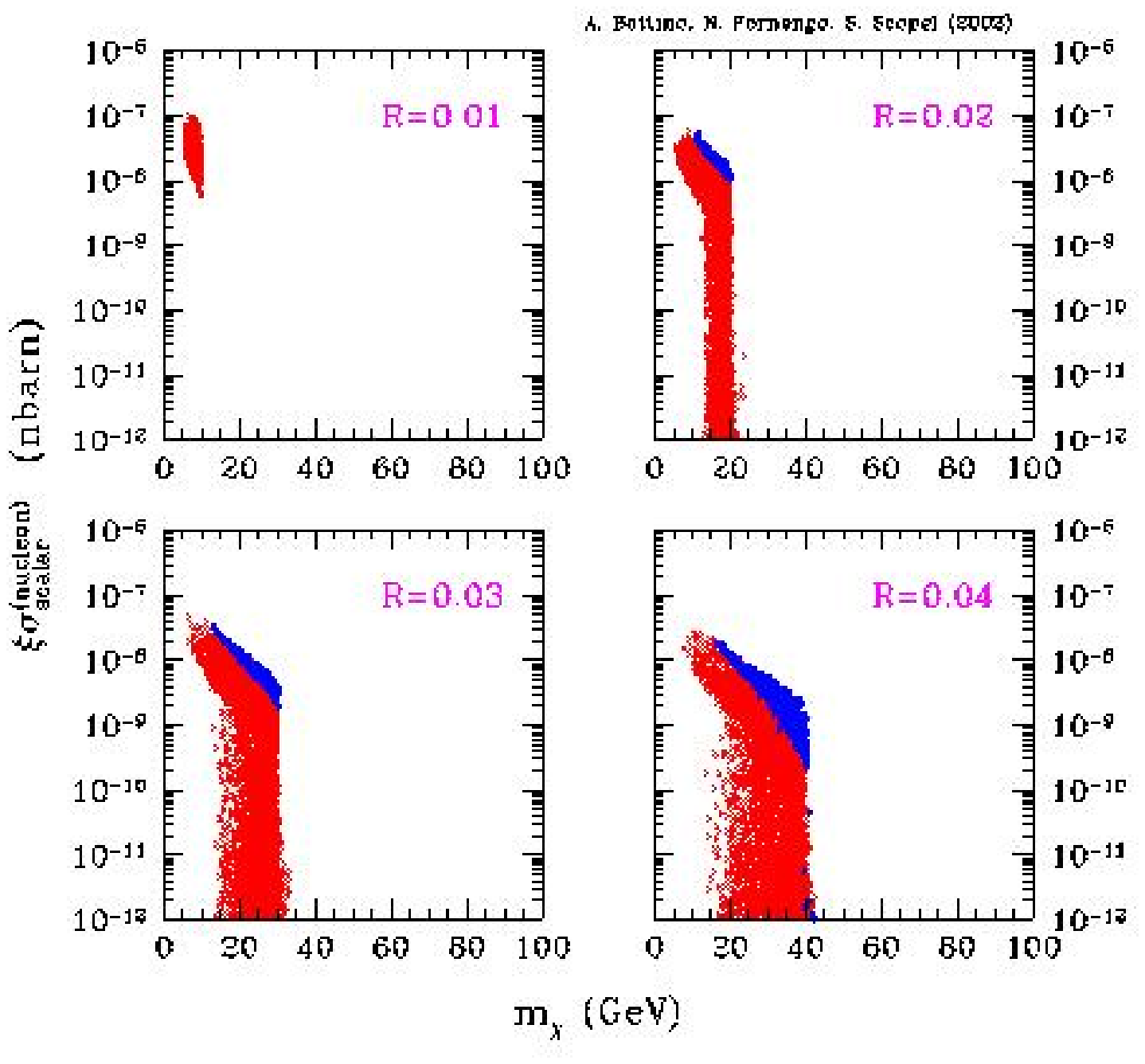}
\caption{\label{fig:1a} Scatter plots of the neutralino--nucleon cross
section $\sigma_{\rm scalar}^{(\rm nucleon)}$ times the rescaling
factor $\xi$ vs. the neutralino mass, for nonuniversal gaugino models
with different values of the gaugino mass ratio $R=M_1/M_2$:
$R=0.01,0.02,0.03,0.04$. Crosses denote configurations with dominant
relic neutralinos ($0.05 \leq \Omega_\chi h^2 \leq 0.3$), while dots
refer to subdominant neutralinos ($\Omega_\chi h^2 < 0.05$).}
\end{figure}

\begin{figure} \centering
\includegraphics{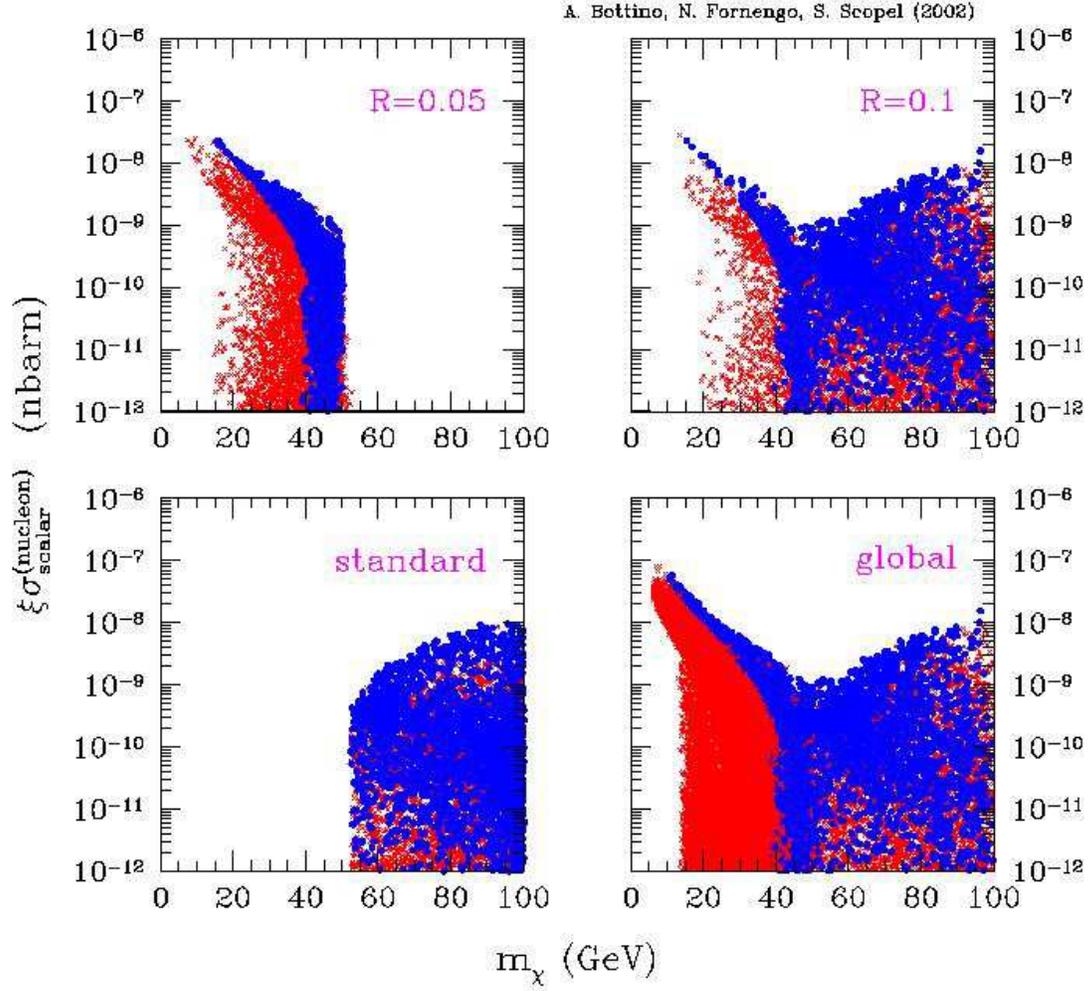}
\caption{\label{fig:1b} The same as in Fig. \ref{fig:1a}, for
$R=0.05,0.1$, for the standard value $R=5/3 \tan^2\theta_W\simeq 0.5$
and for a generic variation of $R$ in the interval 0.01--0.5.}
\end{figure}

\renewcommand{\thefigure}{\arabic{figure}}
\setcounter{figure}{1}
\begin{figure} \centering
\includegraphics{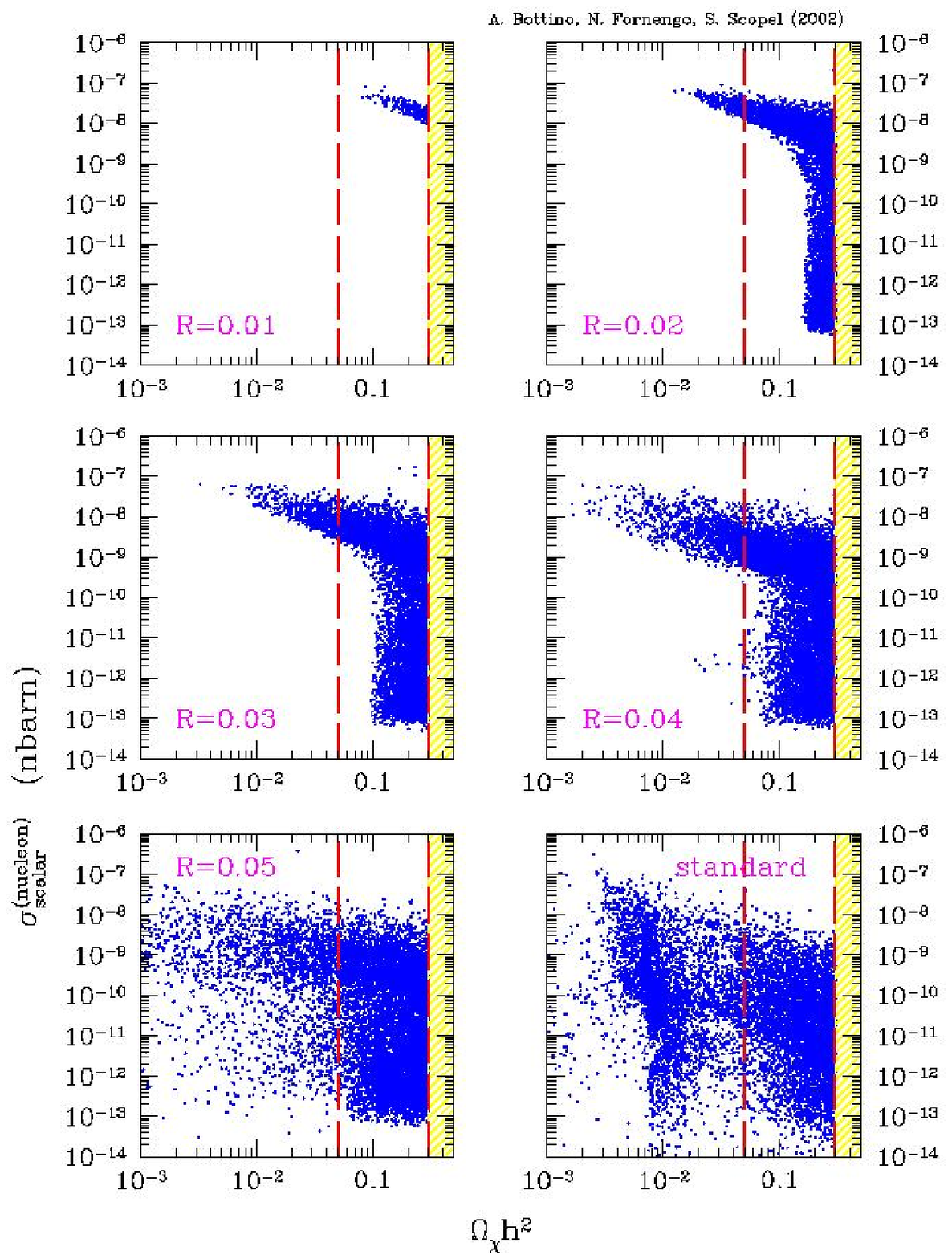}
\caption{\label{fig:2}Scatter plots of the neutralino--nucleon cross
section $\sigma_{\rm scalar}^{(\rm nucleon)}$ vs. the neutralino relic
abundace $\Omega_\chi h^2$, for $R=0.01,0.02,0.03,0.04,0.05$ and for
the standard value $R=5/3 \tan^2\theta_W\simeq 0.5$.}
\end{figure}

\begin{figure} \centering
\includegraphics{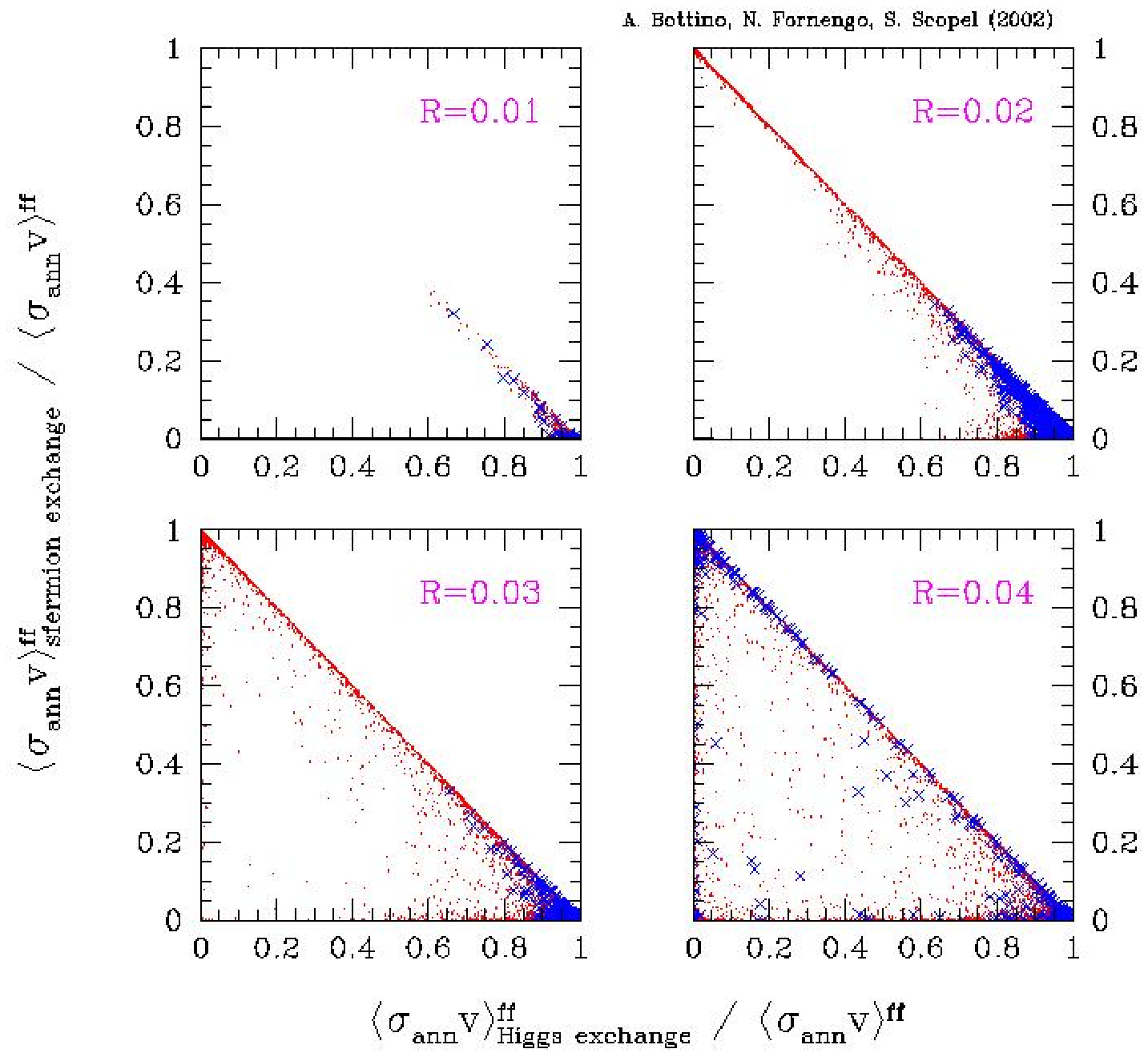}
\caption{\label{fig:3}Scatter plots of the fractional amount of the
neutralino pair--annihilation cross section due to sfermion exchange
vs. Higgs exchange, for $R=0.01,0.02,0.03,0.04$. Crosses denote
configuration for which the neutralino--nucleon scattering cross
section $\sigma_{\rm scalar}^{(\rm nucleon)}$ is larger than $10^{-8}$
nbarn.}
\end{figure}

\begin{figure} \centering
\includegraphics{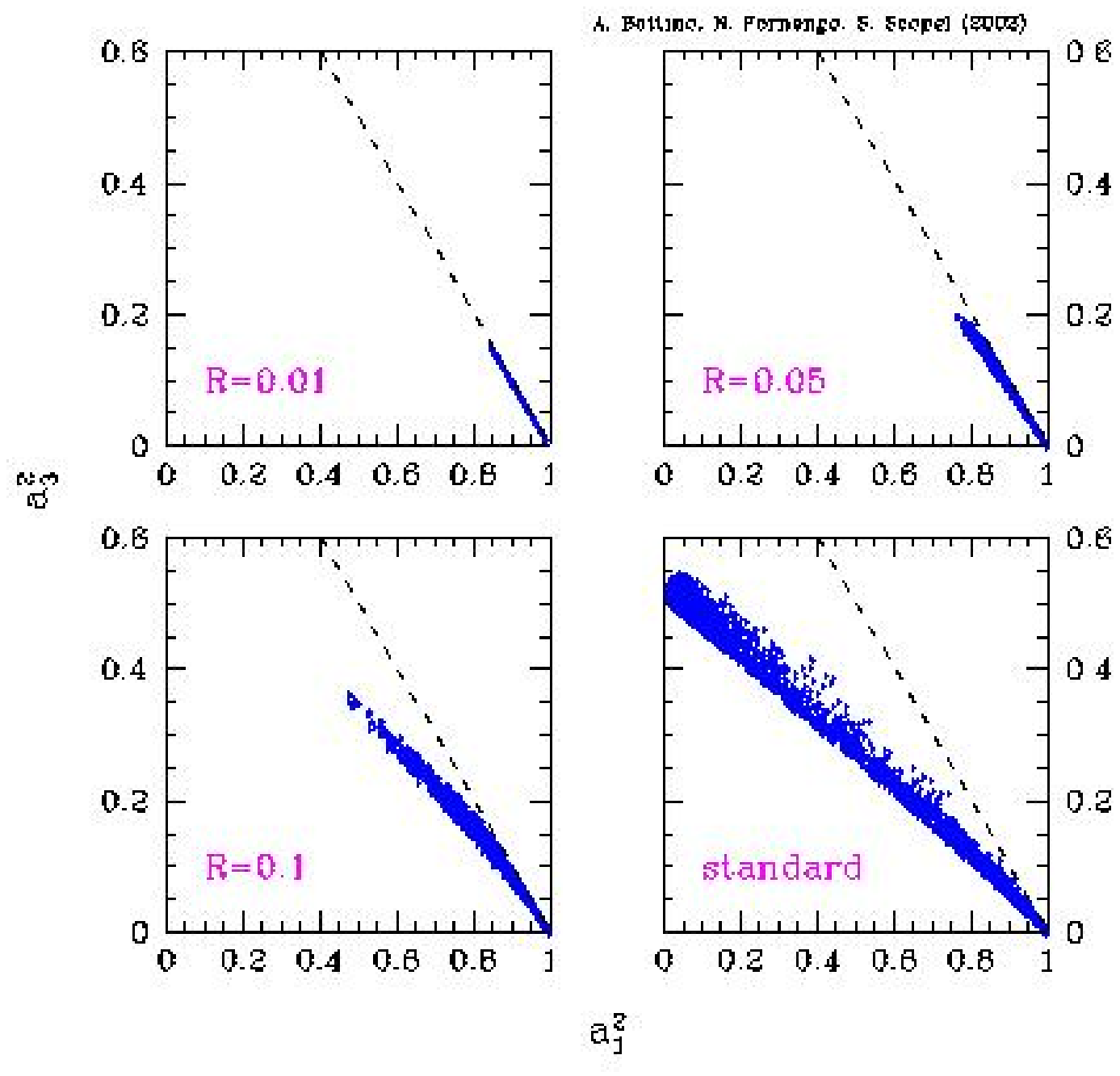}
\caption{\label{fig:4}Scatter plots of the neutralino composition in
terms of $\tilde B$ ($a_1$) and of $\tilde H_1^{\circ}$ ($a_3$) for
$R=0.01,0.05,0.1$ and for the standard value $R=5/3
\tan^2\theta_W\simeq 0.5$. The dashed lines denote the line where
$a_1^2+a_3^2=1$.}
\end{figure}

\end{document}